\documentclass[10pt]{article}

\usepackage{natbib}
\usepackage[margin=2cm]{geometry}

\title{Methods for Exploring Simulation Models}

\author{Juste Raimbault$^{1,2,3}$ et Denise Pumain$^{3}$\\
\medskip\\
$^{1}$ UPS CNRS 3611 ISC-PIF\\
$^{2}$ CASA, UCL\\
$^{3}$ UMR CNRS 8504 G{\'e}ographie-cit{\'e}s
}
\date{}

\begin{document}

\maketitle

\begin{abstract}
Simulation models are an absolute necessity in the human and social sciences, which can only very exceptionally use experimental science methods to construct their knowledge. Models enable the simulation of social processes by replacing the complex interplay of individual and collective actions and reactions with simpler mathematical or computer mechanisms, making it easier to understand the relationships between the causes and the consequences of these interactions and to make predictions. As the formalism of mathematical models offering analytical solutions is often not suitable for representing social complexity~\citep{jensen2018pourquoi}, more and more agent-based computer models are being used. For a long time, the limited computing capacities of computers have hampered programming models that would take into account the interactions between large numbers of geographically located entities (persons or territories). In principle, these models should inform the conditions for the emergence of certain patterns defined at a macro-geographic level from the interactions occurring at a micro-geographic level, in systems whose behaviors are too complex to be understood directly by a human brain. Moreover, it is also necessary to analyze the dynamic behavior of these models with nonlinear feedback effects and verify that they produce plausible results at all stages of their simulation. This essential work of exploring the dynamics of modeled systems remained in its infancy until the late 2010s. Since then, algorithms combining more sophisticated methods, including genetic algorithms and the use of distributed intensive computing, have made it possible to make a significant qualitative leap forward in the exploration and validation of models. The result is an epistemological turn for the human and social sciences, as indicated by the latest applications realized with the help of the OpenMOLE platform presented here.
\end{abstract}

\section{Social sciences and experimentation}

Experimentation has greatly helped build the natural sciences. Experimenting consists in reproducing material, physical, chemical, or biological processes, according to devices designed by researchers to select, often by isolating them, simpler sequences of facts than those operating in a complex reality. The comparison of the results of these manipulations with observed data, in whole or in part unrelated to those used to construct the experimental design, is considered to provide evidence of the veracity or accuracy of the explanatory reasoning underlying the construction of the model. The evidence is supposed to be more or less conclusive depending on the quality of the fit between the model predictions and the observations. However, it is well-known that the accuracy of a model’s predictions is not sufficient to fully validate the adequacy between the explanatory mechanism imagined by the manipulators of the experimental device and the processes at work in the system under study, even if it remains an important step in the construction of models and theories enriched by observations.

In the human and social sciences, the development of experimental devices is very delicate because it faces many practical and ethical obstacles. Ethical and political criticism imposes limits on the manipulation of people and the control of their freedom. The scruples of scientific ontology and ethics (part of what is now called integrity) have certainly not prevented in practice the manipulations, benevolent or not, carried out in historical times by actors with political, cultural, or economic power from making decisions, more or less well informed “scientifically” (see the writings of the “prince’s advisors” across all eras, such as Bodin, Machiavelli, Botero, etc. to name a few of those who have dealt with spatial planning) and to carry out ``experiments'' in forms of governance or technological or cultural innovations whose results have been evaluated, sometimes as beneficial and sometimes as disastrous. Assessing the effectiveness of decisions is even more complicated since there is often no countertest, no alternative scenario, and the actors themselves do not deprive themselves of ``biasing'' the game with their ``self-fulfiling prophecies'', ensuring in advance the result they expect~\citep{rist1970student}. The often deplored difficulty of evaluating public policies is thus exacerbated by the uncertainty of the boundaries between action and context, both in space and time.

Driving change in social life, whatever the scale of the interventions, remains a costly and risky operation, ethically difficult for scientists to accept, and there are few who actually dare to embark on “research-action” projects. During the 1960s, there was a controversy in France between the advocates of “applied geography”, who were familiar with the “terrain” but were sometimes methodologically and politically conservative, and those of “active geography” who were more committed to transforming society. Sometimes, for example, to contribute to the definition of the policy of equilibrium metropolises in France (DATAR\footnote{\emph{D{\'e}l{\'e}gation {\`a} l'Am{\'e}nagement du Territoire et à l'Action R{\'e}gionale} – Inter ministerial delegation of land planning and regional attractiveness.} operation in 1964), the geographers participating in planning studies (including Michel Rochefort in this case) relied, without really daring to say it, on scientific models (in this case Walter Christaller’s central place theory). Today, geographers are more willing to want to assist in decision-making in the most informed way possible according to the state of their knowledge. They often choose to use in silico simulation models, implemented in computers. Computer simulation has thus become a substitute for experimentation. It is no coincidence that, among social scientists, geographers became pioneers in the field: the diversity of the multiple categories of data (landscapes, populations, habitats, etc.) that they manipulate to account for the transformations of the earth’s surface and land interfaces operated by societies, the often large extent of the territories they examine at regional, national, or global levels, explain their need to use computers to organize these masses of information, and understand the dynamics they represent.

\section{Geographical data and computer skills}

We consider a simulation model as a computer program that produces outputs from input data and parameters. In an ideal case, all the following steps are necessary for a robust use of simulation models:
\begin{enumerate}
	\item Identification of the main mechanisms and associated crucial parameters, as well as their range of variation, suggested by their thematic meaning if applicable; identification of indicators to assess the performance or behavior of the model.
	\item Evaluation of stochastic variations: a large number of repetitions for a reasonable number of parameters; establishment of the number of repetitions necessary to achieve a certain level of statistical convergence.
	\item Direct exploration for a first sensitivity analysis; if possible, statistical evaluation of the relationships between parameters and output indicators.
	\item Calibration, algorithmic exploration targeted by the use of specific algorithms, such as \emph{Calibration Profile} \citep{reuillon2015new} and \emph{Pattern Space Exploration} \citep{10.1371/journal.pone.0138212}.
	\item Returns on the model, extension, and new multi-modeling bricks; returns on stylized facts and theory.
	\item Extensive sensitivity analyses, corresponding to experimental methods (under development and integration in the OpenMOLE platform), such as the sensitivity to meta-parameters and initial spatial conditions proposed by \cite{raimbault2019}.
\end{enumerate}

Some steps do not necessarily need to be taken, such as stochasticity assessment in the case of a deterministic model. Similarly, the steps will become more or less important depending on the nature of the question asked: calibration will not be relevant in the case of completely synthetic models, while a systematic exploration of a large number of parameters will not always be necessary in the case of a model that is to be calibrated on data.

The first geographical simulation models were first calculated ``by hand'' in the 1950s. It is no coincidence that these models all deal with stylized facts that reflect the most frequently observed regularities in the organization of social space and which are effects of the ``first law of geography'', summarized as early as 1970 by the American cartographer Waldo Tobler: “everything interacts with everything, but two close things are more likely to come into contact than two distant things”. The attractive power of proximity appears in all social processes of spatial planning, which are constrained by the “obligation of spacing”. This expression was proposed by Henri Reymond in 1971 in a formalization of the problems of geography, which first stated that societies tend to transform the heterogeneous, rough, and discontinuous earth’s surface into an organized space with properties of greater homogeneity and continuity. Thus, regularities emerge because two objects cannot occupy the same place. The assertion that \textit{individuals and societies are most likely to choose to occupy and interact with the nearest locations, both because they are better known and because they save on the costs (physical, monetary, and cultural) of distance travel} is certainly the strongest theoretical proposition in geography. It is found in all spatial configurations that lead to the distinction between a center and a periphery, which appears at all levels of geographical space, from local to global.

The first simulation models in geography therefore first touched on processes for which the choice of the closest, among the places with which one wishes to establish an interaction, is a very dominant anthropological constant. Proximity is preferred in spatial interaction, whether to observe the effects of an innovation before imitating it, according to Torsten Hägerstrand’s spatial theory of innovation diffusion \citep{hagerstrand1957migration} or for the choice of destination places by migrants \citep{hagerstrand1957migration,morrill1962development,morrill1963distribution}. Models based on the proposal, as early as 1954, of the American geographer Edward Ullman to construct geography as a ``science of spatial interactions'', particularly in relation to trade relations, first gave rise mainly to experiments with statistical models, using various forms of ``gravitation models''. These were then integrated into urban models, first static \citep{lowry1964model} and then dynamic \citep{clarke1983dynamics,wilson2014complex,allen1981urban,allen2012cities}.

A later generation of models, which played with the effects of proximity in a more complex way, made extensive use of cellular automata. Spatial autocorrelation measurements, which positively or negatively reflect the attraction or competition effects of proximity, are thus used to test the plausibility of simulated configurations for land-use change, particularly urban growth \citep{white1993cellular,white2015modeling} or the spread of epidemics in the geographical area \citep{cliff2004world}.

However, the development of these models was hindered at an early stage by the computational capabilities of the computers of the time, as the explicit representation of spatial interactions increases as the square of the number of geographical units considered. Thus, statistician Christophe Terrier had to segment his Mirabelle program (\textit{M{\'e}thode Informatis{\'e}e de Recherche et d'Analyse des Bassins par l'Etude des Liaisons Logement-Emploi}) processing data from the 1975 INSEE census before being able to simulate the division of resident populations into employment basins according to commuting between all 36,000 French municipalities \citep{terrier1980mirabelle}. The first simulation model of interactions between cities, designed to reproduce their demographic and economic trajectories influenced by urban functions over a period of 2000 years on a laboratory computer, could only accept a maximum of 400 cities \citep{bura1996multiagent,sanders1997simpop}. The increase in the power of computer calculations has been relatively slow, allowing only the consideration of about a thousand cities in 2007 with the Eurosim model \citep{sanders2007artificial} or the Simpop2 models applied by Anne Bretagnolle to Europe and the United States \citep{bretagnolle2010simulating}. Above all, experimentation with these models has long remained at a low tech stage, requiring a high degree of application in the ``manual'' modification of parameter values, which were rarely directly observable, and which, therefore, had to be estimated on the basis of the plausibility of the model dynamics. However, the equations of urban dynamics models incorporate nonlinear relationships that produce many bifurcations, requiring a very large number of round trips in the procedure for estimating parameter values \citep{sanders2007artificial}. This long work limited the number of simulations from which the estimated parameters could be considered satisfactory, and above all, once the model had been calibrated in this way, there remained a fairly high degree of uncertainty as to the quality of the results obtained.

\section{New generation simulations}

The 2000s and 2010s were to completely change the researchers’ working environment, because the spread of the Internet, then mobile phones, and finally massive data produced by all kinds of digital sensors had rapid and intense feedback effects on the rise in computing capacities, which in turn enabled these disruptive technological innovations. The simulation models were then able to integrate considerable amounts of interactions between localized entities characterized by a wide variety of attributes. Until 15 years ago, \cite{gleyze2005vulnerabilite} was forced to conclude that network analyses for public transport in Paris were ``limited by calculation''. To give just one example of the quantitative leap represented by the increase in computational capacity and its consequences on the increased confidence in models that results from it, we can cite the pioneering work in digital epidemiology carried out by \cite{eubank2004modelling} to simulate, using EpiSims and TRANSIMS models, the daily trajectories, on a transport network, of the movements of one and a half million people between some 180,000 places in the virtual city of Portland, in order to predict the propagation paths of an epidemic based on the probability of interpersonal encounters, in social networks organized in “small worlds”. The epidemic can spread rapidly in a city, while the number of contacts per person remains small (maximum 15, \cite{eliot2006diffusion}).

Simulation platforms have been developed so that as many researchers as possible, even those not specialized in computers, can implement multi-agent models. NetLogo \citep{tisue2004netlogo} is probably the best known; it is generalist, and allows access to multi-agent simulations without the need for in-depth computer knowledge, thanks to its simple programming language and integrated graphical interface builder. Other more specialized platforms, such as GAMA \citep{grignard2013gama}, are immediately built to offer a coupling with geographic information systems. However, confidence in the results from simulation models goes hand in hand with an increase in the size and a number of simulations required, i.e. the scale of numerical experiments. Although platforms such as NetLogo and GAMA integrate basic tools that allow a first step toward such a ``scaling up'', a need for a ``meta-platform'' dedicated to the exploration of simulation results has naturally emerged.

\subsection{A virtual laboratory: the OpenMOLE platform}

Since 2008, the OpenMOLE platform has been designed to explore the dynamics of multi-agent models \citep{reuillon2010declarative,reuillon2013openmole} and has rapidly extended to simulation models more generally. It is the result of the development of a previous software, SimExplorer \citep{amblard2003comprendre,deffuant2003demarche}, which already provided its users with an ergonomic interface for designing experimental plans and gave access to distributed computing. OpenMOLE generalized SimExplorer in its early days, in particular by making it possible to parallel tasks on a massive scale. OpenMOLE\footnote{\texttt{https://openmole.org/}} is a collaborative modeling tool in perpetual evolution: ``A permanent effort of genericity has made it possible to create in a few years a generic, pragmatic, and proven platform for the exploration of complex system models in the form of a dedicated language, text, and graphics, exposing coherent blocks and at the right level of abstraction for the design of numerical experiments distributed on simulation models'' \citep{schmitt2014modelisation}. \cite{reuillon2013openmole} describe the fundamental principles of the platform, while \cite{pumain2017urban} provide a context for the different uses in simulation models for systems of cities.

The procedures (or workflows) proposed in OpenMOLE are described independently of the models and are, therefore, reproducible, reusable, and exchangeable between modelers. A market place is integrated into the software, such as the model library included in NetLogo, and allows users to benefit from exploration scripts that can be used as templates or examples, in a large variety of thematic fields and for all the methods and languages implemented in OpenMOLE (e.g. for calibrating geographical models, analyzing biological networks, and image processing for neuroscience).

OpenMOLE uses a domain-specific language (DSL) \citep{van2002domain} for writing exploration workflows. This practice consists of creating a rating and rules specific to the field of a given problem. It is a kind of programming language dedicated, in this case, to model exploration and associated methods. This one is, of course, not created from scratch but comes as an extension of the underlying language, i.e. \textit{Scala} in the case of OpenMOLE. A reduced number of keywords and primitives make it easy to use even for a user who has no programming knowledge, and the DSL remains very flexible for the advanced user who can use Scala programming. According to \cite{passerat2017reproducible}, the OpenMOLE DSL is one of the key elements of its genericity and accessibility.

One of the OpenMOLE’s main assets is its transparent access to high- performance computing (HPC) environments. The increase in computing resources mentioned above can be physically manifested in different aspects for the modeler: local server, local computing cluster, computing grid (networking of multiple clusters, such as the European EGI computing grid), and cloud computing services. In most cases, their use requires advanced computer skills, which are generally inaccessible to a standard geographer modeler. OpenMOLE includes a library allowing access to the majority of these computing resources, and their mobilization in the DSL is completely transparent to the user. The person can test their script on their own machine and scale it on HPC environments by modifying a single keyword in it.

Since the presentation of the use of DSL and the implementation of scripts is not the objective of this chapter, we refer the reader to the OpenMOLE online documentation for examples of scripts and model exploration. We simply recall the fundamental components of an exploration script: (i) the definition of prototypes, which correspond to the parameters and outputs of the model, which will take different values during the experiment; (ii) the definition of tasks, including the execution of the model but which can also be, for example, pre- or post-processing tasks – tasks covering a very wide variety of languages (Scala, Java, NetLogo, R, Scilab, and native code such as python or C++); (iii) the description of the methods to apply (exploration by sampling, calibration, diversity search, etc.) that will affect the values of the prototypes and launch the evaluation task under consideration (most often the model); (iv) a specification of the data retrieved from the script execution (simulation data is often massive, so a selection of these is crucial); and (v) the definition of the calculation environment on which the method will be launched.

The platform aims to considerably extend the practices of generative social science proposed by \cite{epstein1996growing}, who considered each multi-agent model as an artificial society, generating macroscopic behaviors based on hypotheses about microscopic behaviors. According to \cite{schmitt2014modelisation}, who used the OpenMOLE platform to develop with \cite{rey2015plateforme} the SimpopLocal model to simulate the emergence of a system of cities, the virtual laboratory that this platform represents ``is no longer just the simulation model and the hypotheses it simulates (i.e. the artificial society). It also contains the modeling methods, tools, and procedures adapted to the design and exploration of the model and whose practice provides as much theoretical knowledge and feedback as the design of the model itself. This virtual laboratory is, therefore, all the more similar to a real research laboratory with a bench (the model to be designed and explored), a researcher’s hypotheses (the geographical processes transcribed into the model’s mechanisms), methods (the iterative and intensive computation- assisted modeling method), tools (automated exploration procedures, and any other experimental design incorporated in OpenMOLE), all gathered in a room, the SimProcess modeling platform (an alternative name for the framework in which OpenMOLE is embedded)'' \citep{rey2015plateforme}.

Compared to general protocols such as the one introduced by \cite{grimm2005pattern} to present all the modeling steps, the principles applied in OpenMOLE are novel in terms of their new capabilities to explore the dynamic behavior of simulation models. Two main innovations consist of the systematic use of optimization metaheuristics, mainly genetic algorithms, to quickly test as many combinations of model parameter values as possible, and in the simultaneous sending of simulations to multiple machines on a computing grid, which considerably reduces the duration of experiments, which would otherwise quickly become unacceptable.

The choice of genetic algorithms as optimization heuristics is justified by their effectiveness in multi-objective optimization problems. In addition, the island distribution scheme (populations evolving independently for a certain period of time) is particularly well suited to the distribution of grid calculations, each of the nodes of the grid evolving a sub-population, which is regularly retrieved, merged into the overall population, from which a new sub-population is generated and sent to the node. This type of algorithm could be relatively applicable to stochastic model simulations although a number of problems remain for this type of application \citep{rakshit2017noisy}. According to \cite{rey2015plateforme}, these methods are part of the more global framework of Evolutionary Computation, and the Scala MGO library, developed simultaneously with the platform and which allows evolutionary algorithms to be implemented there, was designed to be easily extended to other heuristics in Evolutionary Computation, leaving completely open the possibility of including new methods in OpenMOLE.

According to R. Reuillon cited by \cite{raimbault2017applied}, OpenMOLE’s philosophy revolves around three axes: the model as the ``black box'' to be explored (i.e. exploration methods ``rotate the model'' without interacting with its code, which is thus somehow ``embedded'' in the platform), the use of advanced exploration methods, and transparent access to intensive computing environments. These different components are strongly interdependent and allow a paradigmatic turn in the use of simulation models: the use of multi-modeling, i.e. a variable structure of the model, as presented in Chapter 4 \citep{cottineau2015modular} and a change in the nature of the questions asked of the model, such as a complete determination of the space for possible model behaviors \citep{10.1371/journal.pone.0138212}, with all this being made possible by the use of intensive computing \citep{schmitt2015half}. The online documentation provides an overview of the methods available in the latest version of the software and their articulation in a standard framework.

To illustrate this general presentation of the OpenMOLE platform and associated methods, we propose developing the example of the SimpopLocal model in the following section, whose genesis has been closely linked to that of the platform, and which has been a candidate for the development and application of various methods.

\subsection{The SimpopLocal experiment: simulation of an emergence in geography}

The SimpopLocal model was designed to represent the emergence of systems of cities, as observed in five or six regions of the world, some 3000 years after the emergence of agricultural practices in settled societies \citep{bairoch1985jericho,marcus2008ancient}. It is indeed a question of explaining the emergence not only of ``the'' city but also of ``systems of cities'', because we know that cities from that time were never isolated but already organized in networks in the territory of each of these ancient ``civilizations''. The most recent publications by archeologists insist on a certain continuity of the processes that have led to the sedentarization of hunter-gatherer populations, grouped into hamlets and villages, and then to the emergence of cities in some of these regions. The development of agriculture has been accompanied by a considerable increase in population densities and the size of human groups in these regions (from 0.1 people per km2 to 10, a factor of 100 between the two orders of magnitude), as well as by a greater complexity of political organization and the social division of labor. This very slow process of resource accumulation and population concentration is carried out according to a series of processes involving substantial feedback, with many fluctuations in growth due to frequent adverse events such as natural disasters or predations from neighboring groups. Due to the slowness of the transformations and their frequent interruptions, archeologists sometimes now contest the name ``Neolithic revolution'', which was proposed by Gordon Childe in 1942 \citep[p. 159]{demoule2018histoire}.

However, geographers continue to identify the emergence of cities as an emergence, a ``bifurcation'' for two main reasons: on the one hand, it has not occurred systematically in all regions where agriculture has been practiced, where two regimes for the evolution of settlement systems are possible and historically viable (purely agricultural and village regions have been able to function for several centuries and now remain residual in some forests or on Pacific islands, for example), and the territorial regime operating with cities is indeed a specific ``attractor'' in the dynamics of old settlement systems; on the other hand, the evolutionary trajectory that sees cities emerge reflects a significant qualitative change (an emergence) with a significant increase in the diversity of social functions associated with habitats and also a considerable expansion in the scale of spatial interactions: the trade that takes place over a longer distance thus allows cities to be less dependent on a “site” of local resources as it is the case for agricultural villages and to develop the assets of a geographical ``situation'' exploiting the wealth of a network of increasingly distant sites \citep{reymond1971pour}.

The SimpopLocal model aims to reproduce this remarkable aspect of the dynamics of settlement systems, which invariably produces an amplification of the hierarchical differentiation between habitats, defined in the literature as a major stylized fact: already in any settlement system, in any place, and at any time in history or prehistory, the distribution of the sizes of inhabited places (measured by population or spatial extent, or even the diversity of functional artefacts) is statistically very asymmetric, comprising many very small agglomerations and only a few very large agglomerations with a fairly regular distribution of the Zipf’s law or lognormal type \citep{fletcher1986settlement,liu1996settlement}. This hierarchical distribution is a structural property (size order of entities) at the macroscopic level that is particularly persistent over time, regardless of local fluctuations at the entity level. The SimpopLocal model is designed to test the hypothesis set out in the evolutionary theory of urban systems \citep{pumain1997pour}, which explains this structural characteristic by a process of urban growth on average proportional to the size acquired, and its amplification by the creation of multiple technological and societal innovations producing the increase and diversification of wealth that spreads among places connected by all kinds of exchanges.

The SimpopLocal model is first of all based on the statistical model, which is an excellent first approximation of the evolution of populations in a system of cities by simulating urban growth as a simple stochastic process that varies the size of each city in a manner proportional to its size and leads to a lognormal distribution of urban populations \citep{gibrat1931inegalites}. The high quality of this basic statistical model is that it uses the size already acquired, which expresses both the accumulated wealth and the attraction and resilience of the inhabited place, as an “explanation” for the growth. In a way, it is a model according to the concept of ``endogenous growth'' of economists \citep{aghion1998endogenous}. However, SimpopLocal is designed, like the previous models of the ``family'' of Simpop multi-agent models \citep{bura1996multiagent,sanders2007artificial,pumain2008socio}, to compensate for the insufficient capacity of Gibrat’s model to predict the everywhere observed trend of higher than expected growth in the largest cities at the top of networks \citep{moriconi1993urbanisation} and the exaggerated inequality between city sizes \citep{pumain1997pour,bretagnolle2010simulating}. These deviations from the Gibrat model are related to long-range correlations \citep{rozenfeld2008laws}, caused by spatial interactions. The effect of these amplifies the hierarchical differentiation between the sizes of cities participating in exchanges in an urban system \citep{favaro2011gibrat}. Simpop models reflect this effect by introducing, exogenously to the model and at different times during the simulation, new urban functions that select certain cities or are captured by them in a continuous process of adaptation to these innovations. Compared to the other Simpop models, SimpopLocal introduces two new features: it uses an abstract representation of successive waves of innovation and brings them all together in a single ``innovation'' object. A second originality consists of making endogenous the process of creating innovation by linking it to the size of the inhabited place, supposed to amplify in a nonlinear way the emergence of new technical, social, or cultural forms (with a probability of creation varying as the square of the populations in presence or in relation). This more parsimonious version of the model construction considerably reduces the number of parameters and, therefore, allows for more systematic exploration and evaluation.

\subsection{Implementation of SimpopLocal, from NetLogo to OpenMOLE}

SimpopLocal was initially developed with the NetLogo language and then redeveloped with the Scala programming language. The simulation with NetLogo benefited from the facilities of the interface, which allows numerically and graphically following the modifications generated on the macroscopic variables that summarize the state of the system, but very quickly showed its limits in terms of experimentation. The manual method of adjusting parameter values made it difficult to avoid the ``runaway'' of urban growth leading to increases in city size that was far too huge for the historical period being simulated. The reprogramming in Scala and then the transfer to the OpenMOLE platform allowed a more precise and complete exploration of the model’s behaviors.

The model represents the evolution of settlement units dispersed in an area large enough to accommodate a few thousand inhabitants but of a surface limited enough to ensure a possible connection between the inhabited places according to the transportation means available at the time (e.g. it could be former Mesopotamia or ancient Mesoamerica). The simulation space is composed of about a hundred inhabited places. Each site is considered a fixed agent and is described by three attributes: the location of its permanent habitat $(x,y)$, the size of its population $P$, and the resources available in its local environment. The amount of available resources $R$ is quantified in units of inhabitants and can be understood as the carrying capacity of the local environment to support a population, which varies according to the skills that the local population has acquired for using resources, through innovations they have created or received from other inhabited places. However, resource exploitation is done locally and the sharing or exchange of goods or people is not explicitly represented in the model. Each new innovation created or acquired by an inhabited place develops its operating skills. The innovation entity is understood here as a large socially accepted abstract invention, which could represent a technical invention, a discovery, a social organization, new habits, or practices, and so on. Each acquisition of innovation by an inhabited place brings the possibility of exceeding its capacity thresholds and, consequently, authorizes demographic growth.

The model was designed to be as parsimonious as possible, minimizing the number of agent attributes (which are inhabited places) and the parameters that control their evolution. The average orders of magnitude indicated by the archeologists’ work were used directly to set the length of the transition period between an agrarian settlement system and an urban settlement system at about 4,000 years, to estimate an average annual population change rate of about 0.02\% per year and to estimate that the size of the largest inhabited place in the system would increase from about 100 to about 10,000 inhabitants. However, the values of five other parameters could not be estimated from the literature and had to be deduced from the simulation experiments. These are the probability of creating an innovation by interaction between two people in the same place, the probability of spreading an innovation by interaction between two people in different places, the intensity of the dissuasive effect of distance on these interactions between places, and the impact of an innovation on population growth (through an increase in available resources) and the maximum possible size of an inhabited place (measured in terms of population or available resources) that is part of the logistics growth equation adopted as a generic model of a development that is still very strongly constrained by local resources. The equations that summarize the model and the tables that precisely define the parameters and their action are detailed by \cite{schmitt2014modelisation} and \citep[pp. 21–34]{schmitt2017simpoplocal} in \cite{pumain2017urban}.

An initial value is defined for the population and resources of the inhabited places, and then the network of interaction between them is created. Then, at each step of the simulation, the mechanisms of population growth and innovation diffusion are applied. The impact of innovations on the efficiency of resource extraction is calculated. This loop is iterated until the stop criterion is reached: in this case, after 4,000 steps or when an arbitrary maximum number of innovations have been reached. The evolution of the state of the settlement system at the macro-geographic level is observed by the distribution of the size of the inhabited places, summarized by the slope of the rank-size distribution. As the model includes certain parameters that are probabilities, it is stochastic; thus, the same set of parameter values can give significantly different results. An automated method for varying parameter values and interpreting the results obtained has been gradually developed through collaboration between computer scientists and geographers.

\subsection{Calibration and validation}

The automation of the exploration of the dynamics generated by simulation models with the OpenMOLE platform uses genetic algorithms that systematically carry out variations in parameter values previously performed ``manually'' by the researcher. The distribution of calculations on a grid infrastructure (a network of computers) also makes it possible to carry out this very large number of combinatorial operations by considerably reducing computing time, thanks to the parallel processing of information. However, the implementation of this new form of model experimentation requires the intervention of the thematic researcher, who must select the precise objectives that their model must satisfy. In return for what could be considered as a reductionist operation, further refinement of the exploration method can lead to an increased confidence in the scientific assumptions of the model.

\subsubsection{Calibration as optimization using genetic algorithms}

Calibration is a procedure that seeks to minimize the gap (called \emph{fitness}) between the behavior simulated by the model and the empirically observed behavior, by incrementally varying unknown values of the model parameters. \cite{stonedahl2011genetic} recalled the difficulties of this exploration, which quickly becomes tedious when conducted manually because of the multiple bifurcations occurring in models where most of the mechanisms linking the variables are nonlinear. An exhaustive exploration of the parameter space is not possible because it would require excessive computation times, which would increase exponentially with the number of parameters. As these procedures also produce large quantities of results, they also require the use of appropriate methods to process and visualize the information generated by the simulations. A whole set of software must, therefore, be implemented to enable the researcher to discover the main dynamic schemes associated with variations in the parameters of their model.

This is where appropriate IT procedures can be used, relating the calibration issue to an optimization problem. Genetic algorithms have been used to calibrate multi-agent systems in several fields, including medicine \citep{castiglione2007optimization}, ecology \citep{duboz2010application}, economics \citep{espinosa2012genetic,stonedahl2010evolutionary}, or hydrology \citep{solomatine1999automatic}. Despite the wide use of multi-agent systems in the social sciences, this method has not been applied very often \citep{heppenstall2007genetic,stonedahl2010evolutionary}. This type of numerical experiment requires the definition of quantitative objectives to assess whether the simulation results are compatible with the experts’ expectations. It is also necessary to know how to manage enormous computational load and how to optimize a fitness function that is susceptible to present very large stochastic variations \citep{di2004applying}.

In the case of SimpopLocal, which includes five parameters whose values are unknown (even their orders of magnitude cannot be estimated from empirical data), we had to identify three ``objective functions''. These characterize a simulation result at the macro-geographic level and correspond to stylized facts whose orders of magnitude could be established on the basis of archeological and historical knowledge: the final distribution of city sizes must be lognormal (similar to a Zipf’s law), the maximum size reached by the largest city must be about 10,000 inhabitants, and the simulation duration must be equivalent to 4,000 years.

This obligation to define objective functions could be considered as a strong constraint on the epistemological validity of the model. It seems to contradict the hypothesis of an open evolution for systems of cities. In fact, this intermediate step of calculation represents a compression of knowledge, our \emph{minimum} requirement on the representativeness, and plausibility of the model’s behavior in relation to the possible set of dynamics of cities in a system (at the historical time of the emergence of cities). The result in terms of evaluation of the simulations must make it possible to progress in the knowledge of the intra- and inter-urban interaction processes likely to generate this general dynamics at the macroscopic level of the system, this theoretical reconstitution thus being similar to what physicists call the ``opposite problem''.

A rather wide range of numerical variation is established \emph{a priori} for each of the five parameters. Each set of parameters, combining a value for each of them, is evaluated according to the simulation output it produces. This evaluation measures the proximity between the simulation outputs and the objective functions defined for the model and thus measures the ability of a certain set of parameter values to reproduce the stylized facts that the simulation should best approach. The settings receiving the best evaluations are then used as a basis to generate new parameter sets that are then tested.

\subsubsection{Exploration of the parameter space under objective constraint}

The SimpopLocal model being stochastic, the simulation results vary from one simulation to another for the same parameter settings. Therefore, the evaluation of the parameter settings according to the three objectives must take this variability into account. We verified that about 100 simulations for each set of parameters were sufficient to capture this variability without significantly increasing the duration of the calculation.

Each objective function has a corresponding measurement evaluating the quality of the simulated result. The model’s ability to produce a log-normal distribution is measured by the difference between the simulated distribution and a theoretical log-normal distribution, having the same mean and standard deviation according to a Kolmogorov–Smirnov test. The maximum population objective quantifies the model’s ability to generate larger or smaller cities, and the result of a simulation is tested by calculating the difference between the size of the largest agglomeration and the expected value of 10,000 inhabitants: $\left|(\textrm{population of the largest agglomeration} - 10,000) / 10,000 \right|$. The simulation duration objective quantifies the model’s ability to generate an expected configuration within a historically plausible time frame. The difference between the number of iterations of the simulation and the expected value of 4,000 simulation steps is calculated: $\left| (duration simulation - 4000) / 4000 \right|$. These three error calculations are standardized in order to compare the degree of success of a simulation with each of the three objectives. However, since it is not possible to aggregate the three calculations to produce a single global quality measure, a multi-objective algorithm is needed to determine which simulations are most satisfactory for approaching the desired final configuration. This type of algorithm calculates compromise solutions such that none dominates all others for all objectives. These solutions are called Pareto compromises and together they form what is called a ``Pareto front''.

The use of global exploration methods such as genetic algorithms to calibrate a multi-agent model (and in particular a stochastic multi-agent model) involves a very high computation cost \citep{sharma2006multi}. This type of load is too large to run on local computers, and supercomputers are very expensive and are not readily available in most laboratories. Computer grids offer a solution to solve these computationally intensive problems. However, computing on such a large scale requires orchestrating the execution of tens of thousands of instances of the model on computers distributed around the world. The cumulative probability of local failures and the problem of optimally distributing the workload on the grid make it very difficult for a lay researcher to use it, as mentioned above. It is, among other things, to overcome these difficulties that the OpenMOLE platform was built \citep{reuillon2010declarative,reuillon2013openmole}. This example of the calibration of the SimpopLocal model shows how OpenMOLE helps modelers to bridge the technical and methodological gap between them and high-performance computing.

The grid infrastructure (EGI) allowed us to use such computing power that half a billion model executions could be performed for SimpopLocal calibration, which otherwise would have required about 20 years of computing with a single computer \citep{schmitt2015half}.

\subsubsection{The calibration profile, a great epistemological progress for SHS}

The result of the calibration process only ensures that the model can reproduce the stylized characteristics of the emergence of a city system, with a fairly accurate assessment of the values of the parameters that together contribute to this evolution. However, it says nothing about how often parameter sets produce plausible behaviors, and how each parameter contributes to changing the model’s behavior. For example, it would be interesting to know when certain parameter values prevent the system from achieving plausible behavior and not to limit itself to knowing only one set of ``optimal'' parameter values.

A new method has been developed to represent the sensitivity of the model to variations in a single parameter, independent of variations in all other parameters \citep{reuillon2015new}. By means of a function that calculates a single numerical value describing the calibration quality for the model, the profile algorithm calculates the lowest possible calibration error when the value of a given parameter is set and the others are free. The algorithm calculates this minimum error for the entire variation range of the studied parameter. For each value of a parameter, the algorithm tries to identify the value sets of the other parameters that produce the best fit of the model to the expected data (the smallest possible error). A graph then represents the variations of this optimal adjustment value according to the variations of the studied parameter. The calibration profile thus shows several possible shapes for this curve. When it shows a clear inflection toward the lowest values for calibration error, for a very small range of variation in the values of the parameter under study, it can be concluded that the order of magnitude of the parameter that satisfies the requirements in terms of model behavior has really been identified. If one of these curves remains flat, it indicates that the parameter has no effect on the model’s behavior and can, therefore, be eliminated. Thus, in the case of SimpopLocal, a parameter imagined as the lifetime of an innovation was finally excluded because variations had no effect on the quality of fit of the model, all things being equal with variations in the other parameters \citep{schmitt2014modelisation}. We, therefore, have the opportunity here to assess the extent to which the \textit{mechanisms devised to build the model are not only sufficient but also necessary} to produce the expected behavior. Within the limits of the theoretical framework and the selection of the stylized facts selected, this is the first time that SHS researchers can reach this type of essential scientific conclusion, thanks to a \textit{validation method that is finally effective for multi-agent simulation models}. This marks a huge advancement in the social sciences from an epistemological point of view. Of course, this is a relative certitude that remains always related to the theoretical framework given by the objects, attributes, and mechanisms selected by researchers to be representative of the observed system.

A complementary form of validation of the model could then be imagined if archeological historians tried to recalibrate it with data from their observations. Indeed, the estimated set of parameters contains values that generate the desired dynamics for a fictitious system but are not fixed in absolute terms. They are related to each other, on the one hand, and to the fictitious data entered, on the other. If the latter are modified to make them compatible with a historically observed settlement system, the model’s ability to simulate its development would then be confirmed not only by reconstructing the trajectories of the population evolution of the inhabited places considered but also by maintaining the relative orders of magnitude of the parameters that generate these dynamics.

\section{Other examples of OpenMOLE applications: network–territory interaction models}

In this section, we propose to illustrate the application of OpenMOLE’s exploration methods and intensive computing alongside another thematic issue: that of interactions between networks and territories. This question has fueled many scientific debates, for which most of the problems remain relatively open. For example, the problem of the ``structuring effects of transport infrastructure'' \citep{bonnafous1974methodologies}, presented by \cite{offner1993effets} as a ``scientific myth'' invoked to justify the cost of a new infrastructure by its impact on regional development, not always observed in the medium term, may, according to A. Bretagnolle (in \cite{offner2014effets}), be observed for larger territories and over time, while taking into account local fluctuations in the dynamics of systems of cities. The debate is thus fueled by contradictory results according to the places and periods studied, and also according to the time lags allowed to measure a relationship between a ``cause'' and its ``effect''. The empirical difficulty of extracting general stylized facts, as well as the conceptual difficulty of geographical entities that are, in fact, in circular causal relationships, is bypassed by a model of the co-evolution of transport networks and territories proposed by \cite{raimbault2018modelisation}. The results obtained are closely linked to the use of OpenMOLE and its exploration and calibration algorithms, of which we will give some illustrations.

The application of multi-objective calibration is essential to the experiment with models of systems of cities corresponding to real situations. For example, \cite{raimbault2018indirect} introduces a model of the evolution of a system of cities over time, similar to the model of \cite{favaro2011gibrat}, but focusing on the effect of the physical transport network. The growth rates of cities are determined by the superposition of several effects: (i) endogenous growth captured by a fixed growth rate corresponding to the Gibrat model; (ii) interactions between cities by a gravity model; and (iii) feedback of flows circulating in the network on the evolution of the population of the crossed cities. This model is calibrated in a non-stationary way over time, i.e. over sliding time windows, in order to take the change in the nature of urban dynamics into account, as observed by \cite{bretagnolle2018vers} with, for example, the changes in transport networks, on the French system of cities between 1830 and 2000. To calibrate the model, the simulated populations are compared to the observed populations. At this stage, the use of a multi-objective calibration algorithm (the NSGA2 algorithm implemented in OpenMOLE) is essential. Indeed, the adjustment can be made, for example, on the basis of an average square error over time and for all cities. However, given the disparities in city size linked to the urban hierarchy, it quickly appears that a single-objective optimization on this error will amount to adjusting the size of the largest cities to the detriment of the majority of cities in the system. The addition of a second objective, taken, for example, as an average square error on population logarithms, makes it possible to better consider all cities. An important result of \cite{raimbault2018indirect} is then the emergence of Pareto fronts for these two objectives, for all the time windows considered. This shows that this type of model must be applied by making a compromise between the adjustment of populations for medium-sized cities and populations for larger cities. This result is possible, thanks to the multi-objective optimization by the genetic algorithm of OpenMOLE.

Another example of the application of the platform’s methods, which illustrates its crucial role, is given by the search for co-evolution regimes, i.e. different forms of correlation over time between the development of a network and the growth of a city or region. If the correlation keeps the same sign, it means that the growth of the territory can precede and, therefore, cause that of the network or vice versa, whereas if it varies over time, its changes make it possible to detect circular causalities and, therefore, a co- evolution between network and territory. According to \cite{raimbault2017identification}, the study of the reasons for time-delayed correlation makes it possible to isolate typical regimes of interaction between network variables and territorial variables. More precisely, \cite{raimbault2018modelisation} defines co-evolution as the existence of causal circular relationships, at the level of a set of entities in a certain spatial realm. In the case of networks and territories, network properties must be locally caused by those of the territories and vice versa. Unidirectional causalities from networks to territories then correspond to the ``structuring effects'' mentioned above. This definition of co-evolution makes it possible to capture the ``congruence'' \citep{offner1993effets} between these objects, in a way, their dynamic reciprocal adaptation. It also allows the construction of an operational method proposed by \cite{raimbault2017identification}, which statistically looks for causal links between corresponding variables. In practice, the weak notion of \emph{Granger causality}\footnote{Low causality according to the Granger concept is derived from correlations measured over time series, such causality is assumed when one variable provides results useful for predicting the value of another variable.} is mobilized, allowing flexibility with regard to the necessary data and the temporal and spatial frameworks of estimation. This causality is quantified by the delayed correlations between variations in network variables (such as centralities or accessibility) and variations in territorial variables (such as population, employment, real estate transactions, etc.). The existence of significant maxima at non-zero delays (stronger correlation between the values of one and the growth of the other with a time lag) gives a causal direction (the growth of a network variable precedes and, therefore, leads to the growth of a territory variable or vice versa, for the region and period considered). A typology of these delayed correlation profiles provides what are called ``causality regimes'', including co-evolution regimes where two variables ``territory'' and ``network'' are in reciprocal causality.

The question is then, for a given case study, to identify the regimes present from observed data or data simulated by a model, in particular those that correspond to a co-evolution. The demonstration of the existence of such regimes at the end of a ``co-evolution model'' is not expected \textit{a priori} since the processes included at the microscopic level where the influences are, indeed, reciprocal and do not imply a reciprocal causality at the macroscopic level of the indicators. Indeed, the models considered are complex and reflect an emergence. The delayed correlation method is applied to a macroscopic co-evolution model by \cite{raimbault2018modeling}, which extends the Raimbault model \citep{raimbault2018indirect} by adding rules for the evolution of network link capacities. Direct sampling, which consists of a random draw of a fixed number of values or ``points'' of parameters (e.g. by Latin Hypercube sampling maximizing point spacing), is a first experiment allowed by OpenMOLE to gain an insight into the model’s ability to produce co-evolution. This experiment makes it possible to isolate a number of co- evolution regimes that can potentially be produced by the model: 33 regimes for 729 possible regimes for the variables considered, i.e. 4.5\% (in this case, populations are considered as territory variables, and proximity centrality and accessibility as network variables, which corresponds to six directed pairs of variables, and therefore $3^6=729$ possible regimes, each pair being able to have a positive, negative, or non-existent delayed correlation). Among the 33 isolated co-evolution regimes, there are 19 whose existence could not be intuitively predicted. The existence and variety of these regimes are an important result, showing that it is possible to model a co-evolution in the statistical sense given earlier.

The application of the \emph{Pattern Space Exploration} algorithm \citep{10.1371/journal.pone.0138212}, which aims to achieve diversity in the regimes produced, allows this conclusion to be considerably extended since this algorithm produces 260 co-evolution regimes (35.7\%). This is a typical example where the strong nonlinearity of the outputs considered can lead to partial or even biased conclusions and where the use of a specific method is crucial. The results are made more robust and extensive, thanks to the application of a specific method integrated into the OpenMOLE platform.

This method also allows models to be compared with each other. Thus, \cite{raimbault2018unveiling} applied this method to the SimpopNet model introduced by \cite{schmitt2014modelisation}, which is also a macroscopic co-evolution model and has a large number of similarities with \citep{raimbault2018modeling}'s co-evolution model, particularly in the variables considered, i.e. the calculable output indicators. SimpopNet then produces a smaller number of interaction and co-evolution regimes, confirming, on the one hand, that it is not immediate for a model designed for co-evolution (SimpopNet) to actually bring about co-evolution regimes and suggesting, on the other hand, that its stronger constraints in network evolution rules induce a greater difficulty in producing a diversity of regimes.

\section{Perspectives}

The development of the OpenMOLE platform has made it possible to create an original axis, or even a field of research, one of the remarkable aspects of which is a high level of interdisciplinarity between the human sciences and more technical disciplines such as computer science. According to \cite{banos2017knowledge}, this leads to the production of a broader and deeper knowledge, like the virtuous spiral of \cite{banos2013pour} between discipline and interdisciplinarity. In addition, the single-platform philosophy (mentioned above, through the strong interaction between the three axes of model loading, access to innovative exploration methods, and transparent access to intensive computing environments) opens up many perspectives, both technical and theoretical and methodological and thematic. We give some illustrations in the following, reflecting a present state of possible futures for OpenMOLE.

\subsection{Methods}

The integration of new exploration methods is a key focus of OpenMOLE’s development. For example, the comprehensive resolution of inverse problems \citep{aster2018parameter} is not currently included. The resolution of a reverse problem consists of determining the entire history of a given objective in the output space of the model. Calibration algorithms solve similar problems but do not guarantee the exhaustiveness of the solutions produced, which can be a major problem in the case of equifinality \citep{rey2015plateforme}, i.e. parameter configurations or initial conditions leading by different trajectories to an identical result. A reverse problem heuristic based on Pattern Space Exploration (PSE) mechanisms is currently being developed for integration into OpenMOLE.

The use of Bayesian inference methods is another avenue being explored. Indeed, in the case of highly stochastic models where the attached distributions have non-standard shapes, an estimate of the probability distribution of the parameters can be provided by this type of method. In the case of simulation models, the Approximate Bayesian Computation method \citep{csillery2010approximate} allows, for a set of observed data, the provision of the probability distribution of the parameters that most likely led to these data. It is thus an extended calibration, with a probabilistic produced knowledge allowing uncertainty to be taken into account. A specification of this method proposed by \cite{lenormand2013adaptive}, designed to reduce the number of simulations in the case of models with significant computation time, is also being adapted to parallel computation and integrated into the platform.

We should also mention various methodological disciplines that are also under investigation: (i) the question of high dimensionality quickly becomes a problem in the use of the PSE algorithm since the number of output configurations is potentially subject to the curse of dimensionality, i.e. execution time or size are exponential functions of the number of dimensions – reducing the number of simulated configurations subject to the PSE algorithm by imposing a diversity of selected configurations criterion would solve this problem and take into account a much greater wealth of outputs; (ii) the issue of sensitivity to initial spatial conditions already mentioned  \citep{raimbault2018space} is particularly relevant for geographical models, and a Scala library including synthetic generators of population configurations at different scales is currently being developed, including, for example, the neighborhood generators studied by \cite{raimbault2019generating}; and (iii) the implementation of information criteria for model performance, already mentioned in Chapter 4 and which are a cornerstone of multi-modeling approaches, is also under study, as the POMIC criterion proposed by \cite{piou2009proposing}.

\subsection{Tools}

During its development, OpenMOLE was always at the forefront in terms of the tools used and developed. The choice of the Scala language to replace Java from the first versions of OpenMOLE is an innovative technological choice, made particularly relevant by the functional programming possibilities but also the object programming it provides, making it a more powerful language in terms of flexibility than other functional languages, such as Haskell \citep{oliveira2010scala}. However, the underlying Java infrastructure is retained, allowing high portability to any operating system and any type of hardware, which is very important for distributing calculations over heterogeneous grid nodes. Properties such as line mixing make Scala particularly relevant for multi-modeling \citep{odersky2005scalable}. In addition, properties such as implicit conversions or case classes make Scala ergonomic for DSL development \citep{sloane2008experiences}, which, as already mentioned, is an essential aspect of OpenMOLE.

Questions of program loading, and of models by extension, remain an active field of research, particularly in relation to reproducibility. The \emph{docker} program, which uses \emph{containers}, allows the embedding of an identical execution environment regardless of the operating system and hardware. \cite{hung2016guidock} propose coupling dockers with a graphical interface for scientific reproducibility. Similar programs such as Singularity are specifically dedicated to the reproducibility of HPC experiments \citep{kurtzer2017singularity}. The core of OpenMOLE’s embedding strategy is not based on such a program (Singularity), yet some tasks based on running programs with a complicated environment are embedded in OpenMOLE by a task using docker (e.g. for the R language that requires the installation of a complete R environment). An improvement in the integration of dockers in OpenMOLE is an active and important research axis for the future extension of the genericity of the programs that can be embedded in the platform. OpenMOLE would thus be at the forefront of technical research in terms of scientific reproducibility. Similarly, the question of the scalability of experiments (i.e. the ability of programs to deal with problems of increasing magnitude) is at the heart of the platform’s philosophy, and research is being conducted to, for example, automate the deployment of multiple instances of OpenMOLE on a cluster and to facilitate its use within communities of thematicians.

\section{Conclusion}

The exploration of simulation models continued in geography through initiatives such as the development of the OpenMOLE platform. This development has been carried out in a highly interdisciplinary and reciprocal framework, thus constituting a win–win relationship between computer scientists and geographers \cite{pumaincomplexity}, but also through an unprecedented integration of knowledge domains  \citep{raimbault2017applied}, i.e. empirical, theoretical, and modeling knowledge, but also tools and methods, which strongly interact with one another in each of these domains. The OpenMOLE adventure, and its branch related to geography within the framework of the ERC GeoDiverCity, testifies to a new way of producing geographical knowledge, making it possible to produce scientific evidence in the social sciences. It thus gives weight to geography and, more broadly, to the human and social sciences, in the face of hard sciences such as physics, which claim a monopoly on the identification of evidence relating to the functioning of social systems \citep{dupuy2015sciences}. It now remains to promote this research position, as well as the tools and methods that make it possible, in cooperation with the new emerging disciplines of \emph{City Science} and \emph{Urban Analytics} described by \cite{batty2019urban}, within the field of Theoretical and Quantitative Geography.


\begin{thebibliography}{}

\bibitem[Aghion et~al., 1998]{aghion1998endogenous}
Aghion, P., Howitt, P.~W., Howitt, P., Brant-Collett, M.,
  Garc{\'\i}a-Pe{\~n}alosa, C., et~al. (1998).
\newblock {\em Endogenous growth theory}.
\newblock MIT press.

\bibitem[Allen, 2012]{allen2012cities}
Allen, P.~M. (2012).
\newblock {\em Cities and regions as self-organizing systems: models of
  complexity}.
\newblock Routledge.

\bibitem[Allen and Sanglier, 1981]{allen1981urban}
Allen, P.~M. and Sanglier, M. (1981).
\newblock Urban evolution, self-organization, and decisionmaking.
\newblock {\em Environment and Planning A}, 13(2):167--183.

\bibitem[Amblard, 2003]{amblard2003comprendre}
Amblard, F. (2003).
\newblock {\em Comprendre le fonctionnement de simulations sociales
  individus-centr{\'e}es: Application {\`a} des mod{\`e}les de dynamiques
  d'opinions}.
\newblock PhD thesis, Universit{\'e} Blaise Pascal-Clermont-Ferrand II.

\bibitem[Aster et~al., 2018]{aster2018parameter}
Aster, R.~C., Borchers, B., and Thurber, C.~H. (2018).
\newblock {\em Parameter estimation and inverse problems}.
\newblock Elsevier.

\bibitem[Bairoch, 1985]{bairoch1985jericho}
Bairoch, P. (1985).
\newblock {\em De j{\'e}richo {\`a} Mexico}, volume~4.
\newblock Gallimard.

\bibitem[Banos, 2013]{banos2013pour}
Banos, A. (2013).
\newblock {\em Pour des pratiques de mod{\'e}lisation et de simulation
  lib{\'e}r{\'e}es en g{\'e}ographie et SHS}.
\newblock PhD thesis, Universit{\'e} Paris 1 Panth{\'e}on Sorbonne.

\bibitem[Banos, 2017]{banos2017knowledge}
Banos, A. (2017).
\newblock Knowledge accelerator' in geography and social sciences: Further and
  faster, but also deeper and wider.
\newblock In {\em Urban Dynamics and Simulation Models}, pages 119--123.
  Springer.

\bibitem[Batty, 2019]{batty2019urban}
Batty, M. (2019).
\newblock Urban analytics defined.

\bibitem[Bonnafous and Plassard, 1974]{bonnafous1974methodologies}
Bonnafous, A. and Plassard, F. (1974).
\newblock Les m{\'e}thodologies usuelles de l'{\'e}tude des effets structurants
  de l'offre de transport.
\newblock {\em Revue {\'e}conomique}, 25(25):208--232.

\bibitem[Bretagnolle and Franc, 2018]{bretagnolle2018vers}
Bretagnolle, A. and Franc, A. (2018).
\newblock Vers des syst{\`e}mes de villes int{\'e}gr{\'e}s.
\newblock In {\em Sanders, L., ed. Peupler la terre: De la pr{\'e}histoire à
  l'{\`e}re des m{\'e}tropoles. Presses universitaires François-Rabelais.}

\bibitem[Bretagnolle and Pumain, 2010]{bretagnolle2010simulating}
Bretagnolle, A. and Pumain, D. (2010).
\newblock Simulating urban networks through multiscalar space-time dynamics:
  Europe and the united states, 17th-20th centuries.
\newblock {\em Urban Studies}, 47(13):2819--2839.

\bibitem[Bura et~al., 1996]{bura1996multiagent}
Bura, S., Gu{\'e}rin-Pace, F., Mathian, H., Pumain, D., and Sanders, L. (1996).
\newblock Multiagent systems and the dynamics of a settlement system.
\newblock {\em Geographical analysis}, 28(2):161--178.

\bibitem[Castiglione et~al., 2007]{castiglione2007optimization}
Castiglione, F., Pappalardo, F., Bernaschi, M., and Motta, S. (2007).
\newblock Optimization of haart with genetic algorithms and agent-based models
  of hiv infection.
\newblock {\em Bioinformatics}, 23(24):3350--3355.

\bibitem[Ch{\'e}rel et~al., 2015]{10.1371/journal.pone.0138212}
Ch{\'e}rel, G., Cottineau, C., and Reuillon, R. (2015).
\newblock Beyond corroboration: Strengthening model validation by looking for
  unexpected patterns.
\newblock {\em PLoS ONE}, 10(9):e0138212.

\bibitem[Clarke and Wilson, 1983]{clarke1983dynamics}
Clarke, M. and Wilson, A.~G. (1983).
\newblock The dynamics of urban spatial structure: progress and problems.
\newblock {\em Journal of regional science}, 23(1):1--18.

\bibitem[Cliff et~al., 2004]{cliff2004world}
Cliff, A.~D., Haggett, P., Smallman-Raynor, M., et~al. (2004).
\newblock {\em World atlas of epidemic diseases}.
\newblock Arnold London.

\bibitem[Cottineau et~al., 2015]{cottineau2015modular}
Cottineau, C., Reuillon, R., Chapron, P., Rey-Coyrehourcq, S., and Pumain, D.
  (2015).
\newblock A modular modelling framework for hypotheses testing in the
  simulation of urbanisation.
\newblock {\em Systems}, 3(4):348--377.

\bibitem[Csill{\'e}ry et~al., 2010]{csillery2010approximate}
Csill{\'e}ry, K., Blum, M.~G., Gaggiotti, O.~E., and Fran{\c{c}}ois, O. (2010).
\newblock Approximate bayesian computation (abc) in practice.
\newblock {\em Trends in ecology \& evolution}, 25(7):410--418.

\bibitem[Deffuant et~al., 2003]{deffuant2003demarche}
Deffuant, G., Amblard, F., Duboz, R., and Ramat, E. (2003).
\newblock Une d{\'e}marche exp{\'e}rimentale pour la simulation
  individus-centr{\'e}e.
\newblock {\em Actes des 10{\`e}mes Journ{\'e}es de Rochebrune, rencontres
  interdisciplinaires sur les syst{\`e}mes complexes naturels et artificiels,
  statut {\'e}pist{\'e}mologique de la simulation}, pages 45--64.

\bibitem[Demoule et~al., 2018]{demoule2018histoire}
Demoule, J.-P., Garcia, D., and Schnapp, A. (2018).
\newblock {\em Une histoire des civilisations}.
\newblock Paris, La Découverte, INRAP.

\bibitem[Di~Pietro et~al., 2004]{di2004applying}
Di~Pietro, A., While, L., and Barone, L. (2004).
\newblock Applying evolutionary algorithms to problems with noisy,
  time-consuming fitness functions.
\newblock In {\em Proceedings of the 2004 Congress on Evolutionary Computation
  (IEEE Cat. No. 04TH8753)}, volume~2, pages 1254--1261. IEEE.

\bibitem[Duboz et~al., 2010]{duboz2010application}
Duboz, R., Versmisse, D., Travers, M., Ramat, E., and Shin, Y.-J. (2010).
\newblock Application of an evolutionary algorithm to the inverse parameter
  estimation of an individual-based model.
\newblock {\em Ecological modelling}, 221(5):840--849.

\bibitem[Dupuy and Benguigui, 2015]{dupuy2015sciences}
Dupuy, G. and Benguigui, L.~G. (2015).
\newblock Sciences urbaines: interdisciplinarit{\'e}s passive, na{\"\i}ve,
  transitive, offensive.
\newblock {\em M{\'e}tropoles}, (16).

\bibitem[Eliot and Daud{\'e}, 2006]{eliot2006diffusion}
Eliot, E. and Daud{\'e}, {\'E}. (2006).
\newblock Diffusion des {\'e}pid{\'e}mies et complexit{\'e}s g{\'e}ographiques.
  perspectives conceptuelles et m{\'e}thodologiques.
\newblock {\em Espace populations soci{\'e}t{\'e}s. Space populations
  societies}, (2006/2-3):403--416.

\bibitem[Epstein and Axtell, 1996]{epstein1996growing}
Epstein, J.~M. and Axtell, R. (1996).
\newblock {\em Growing artificial societies: social science from the bottom
  up}.
\newblock Brookings Institution Press.

\bibitem[Espinosa, 2012]{espinosa2012genetic}
Espinosa, O.~B. (2012).
\newblock A genetic algorithm for the calibration of a micro-simulation model.
\newblock {\em arXiv preprint arXiv:1201.3456}.

\bibitem[Eubank et~al., 2004]{eubank2004modelling}
Eubank, S., Guclu, H., Kumar, V.~A., Marathe, M.~V., Srinivasan, A., Toroczkai,
  Z., and Wang, N. (2004).
\newblock Modelling disease outbreaks in realistic urban social networks.
\newblock {\em Nature}, 429(6988):180.

\bibitem[Favaro and Pumain, 2011]{favaro2011gibrat}
Favaro, J.-M. and Pumain, D. (2011).
\newblock Gibrat revisited: An urban growth model incorporating spatial
  interaction and innovation cycles. 再评吉尔布瑞特:
  顾及空间相互作用和创新周期的城市增长模型.
\newblock {\em Geographical Analysis}, 43(3):261--286.

\bibitem[Fletcher, 1986]{fletcher1986settlement}
Fletcher, R. (1986).
\newblock Settlement archaeology: World-wide comparisons.
\newblock {\em World Archaeology}, 18(1):59--83.

\bibitem[Gibrat, 1931]{gibrat1931inegalites}
Gibrat, R. (1931).
\newblock {\em Les in{\'e}galit{\'e}s {\'e}conomiques: Applications... d'une
  loi nouvelle, la loi de l'effet proportionnel}.
\newblock PhD thesis, Recueil Sirey.

\bibitem[Gleyze, 2005]{gleyze2005vulnerabilite}
Gleyze, J.-F. (2005).
\newblock {\em La vuln{\'e}rabilit{\'e} structurelle des r{\'e}seaux de
  transport dans un contexte de risques}.
\newblock PhD thesis, Universit{\'e} Paris-Diderot-Paris VII.

\bibitem[Grignard et~al., 2013]{grignard2013gama}
Grignard, A., Taillandier, P., Gaudou, B., Vo, D.~A., Huynh, N.~Q., and
  Drogoul, A. (2013).
\newblock Gama 1.6: Advancing the art of complex agent-based modeling and
  simulation.
\newblock In {\em International Conference on Principles and Practice of
  Multi-Agent Systems}, pages 117--131. Springer.

\bibitem[Grimm et~al., 2005]{grimm2005pattern}
Grimm, V., Revilla, E., Berger, U., Jeltsch, F., Mooij, W.~M., Railsback,
  S.~F., Thulke, H.-H., Weiner, J., Wiegand, T., and DeAngelis, D.~L. (2005).
\newblock Pattern-oriented modeling of agent-based complex systems: lessons
  from ecology.
\newblock {\em science}, 310(5750):987--991.

\bibitem[Hagerstrand, 1957]{hagerstrand1957migration}
Hagerstrand, T. (1957).
\newblock Migration and area: migration in sweden: Lund studies in geography,
  series b.
\newblock {\em Human Geography}, 13.

\bibitem[Heppenstall et~al., 2007]{heppenstall2007genetic}
Heppenstall, A.~J., Evans, A.~J., and Birkin, M.~H. (2007).
\newblock Genetic algorithm optimisation of an agent-based model for simulating
  a retail market.
\newblock {\em Environment and Planning B: Planning and Design},
  34(6):1051--1070.

\bibitem[Hung et~al., 2016]{hung2016guidock}
Hung, L.-H., Kristiyanto, D., Lee, S.~B., and Yeung, K.~Y. (2016).
\newblock Guidock: using docker containers with a common graphics user
  interface to address the reproducibility of research.
\newblock {\em PloS one}, 11(4):e0152686.

\bibitem[Jensen, 2018]{jensen2018pourquoi}
Jensen, P. (2018).
\newblock {\em Pourquoi la soci{\'e}t{\'e} ne se laisse pas mettre en
  {\'e}quations}.
\newblock Le Seuil.

\bibitem[Kurtzer et~al., 2017]{kurtzer2017singularity}
Kurtzer, G.~M., Sochat, V., and Bauer, M.~W. (2017).
\newblock Singularity: Scientific containers for mobility of compute.
\newblock {\em PloS one}, 12(5):e0177459.

\bibitem[Lenormand et~al., 2013]{lenormand2013adaptive}
Lenormand, M., Jabot, F., and Deffuant, G. (2013).
\newblock Adaptive approximate bayesian computation for complex models.
\newblock {\em Computational Statistics}, 28(6):2777--2796.

\bibitem[Liu, 1996]{liu1996settlement}
Liu, L. (1996).
\newblock Settlement patterns, chiefdom variability, and the development of
  early states in north china.
\newblock {\em Journal of anthropological archaeology}, 15(3):237--288.

\bibitem[Lowry, 1964]{lowry1964model}
Lowry, I.~S. (1964).
\newblock A model of metropolis.

\bibitem[Marcus and Sabloff, 2008]{marcus2008ancient}
Marcus, J. and Sabloff, J.~A. (2008).
\newblock {\em The ancient city: New perspectives on urbanism in the old and
  new world}.
\newblock School for Advanced Research on the.

\bibitem[Moriconi-Ebrard, 1993]{moriconi1993urbanisation}
Moriconi-Ebrard, F. (1993).
\newblock {\em L'urbanisation du monde de 1950 {\`a} 1990 d'apr{\`e}s une
  d{\'e}finition harmonis{\'e}e des agglom{\'e}rations urbaines}.
\newblock PhD thesis, Paris 1.

\bibitem[Morrill, 1962]{morrill1962development}
Morrill, R.~L. (1962).
\newblock The development of models of migration and the role of electronic
  processing machines.
\newblock {\em J. Sutter. Entret. Monaco Sc. Hum. I}, page 213.

\bibitem[Morrill, 1963]{morrill1963distribution}
Morrill, R.~L. (1963).
\newblock The distribution of migration distances.
\newblock In {\em Papers of the Regional Science Association}, volume~11, pages
  73--84. Springer.

\bibitem[Odersky and Zenger, 2005]{odersky2005scalable}
Odersky, M. and Zenger, M. (2005).
\newblock Scalable component abstractions.
\newblock In {\em ACM Sigplan Notices}, volume~40, pages 41--57. ACM.

\bibitem[Offner, 1993]{offner1993effets}
Offner, J.-M. (1993).
\newblock Les ``effets structurants'' du transport: mythe politique,
  mystification scientifique.
\newblock {\em L'espace g{\'e}ographique}, pages 233--242.

\bibitem[Offner et~al., 2014]{offner2014effets}
Offner, J.-M., Beaucire, F., Delaplace, M., Fr{\'e}mont, A., Ninot, O.,
  Bretagnolle, A., and Pumain, D. (2014).
\newblock Les effets structurants des infrastructures de transport.
\newblock {\em Espace Geographique}, 43(1):p--51.

\bibitem[Oliveira and Gibbons, 2010]{oliveira2010scala}
Oliveira, B.~C. and Gibbons, J. (2010).
\newblock Scala for generic programmers: comparing haskell and scala support
  for generic programming.
\newblock {\em Journal of functional programming}, 20(3-4):303--352.

\bibitem[Passerat-Palmbach et~al., 2017]{passerat2017reproducible}
Passerat-Palmbach, J., Reuillon, R., Leclaire, M., Makropoulos, A., Robinson,
  E.~C., Parisot, S., and Rueckert, D. (2017).
\newblock Reproducible large-scale neuroimaging studies with the openmole
  workflow management system.
\newblock {\em Frontiers in neuroinformatics}, 11:21.

\bibitem[Piou et~al., 2009]{piou2009proposing}
Piou, C., Berger, U., and Grimm, V. (2009).
\newblock Proposing an information criterion for individual-based models
  developed in a pattern-oriented modelling framework.
\newblock {\em Ecological Modelling}, 220(17):1957--1967.

\bibitem[Pumain, 1997]{pumain1997pour}
Pumain, D. (1997).
\newblock Pour une th{\'e}orie {\'e}volutive des villes.
\newblock {\em L'Espace g{\'e}ographique}, pages 119--134.

\bibitem[Pumain, 2008]{pumain2008socio}
Pumain, D. (2008).
\newblock The socio-spatial dynamics of systems of cities and innovation
  processes: a multi-level model.
\newblock In {\em The Dynamics of Complex Urban Systems}, pages 373--389.
  Springer.

\bibitem[Pumain, 2019]{pumaincomplexity}
Pumain, D. (2019).
\newblock Complexity in geography.
\newblock {\em Geographical Modeling}, pages 1--30.

\bibitem[Pumain and Reuillon, 2017]{pumain2017urban}
Pumain, D. and Reuillon, R. (2017).
\newblock {\em Urban dynamics and simulation models}.
\newblock Springer.

\bibitem[Raimbault, 2017a]{raimbault2017applied}
Raimbault, J. (2017a).
\newblock An applied knowledge framework to study complex systems.
\newblock In {\em Complex Systems Design \& Management}, pages 31--45.

\bibitem[Raimbault, 2017b]{raimbault2017identification}
Raimbault, J. (2017b).
\newblock Identification de causalités dans des données spatio-temporelles.
\newblock In {\em Spatial Analysis and GEOmatics 2017}.

\bibitem[Raimbault, 2018a]{raimbault2018indirect}
Raimbault, J. (2018a).
\newblock Indirect evidence of network effects in a system of cities.
\newblock {\em Environment and Planning B: Urban Analytics and City Science},
  page 2399808318774335.

\bibitem[Raimbault, 2018b]{raimbault2018modelisation}
Raimbault, J. (2018b).
\newblock Mod{\'e}lisation des interactions entre r{\'e}seaux de transport et
  territoires: une approche par la co-{\'e}volution.
\newblock In {\em JJC Pacte-Citeres-Les capacit{\'e}s transformatives des
  r{\'e}seaux dans la fabrique des territoires}.

\bibitem[Raimbault, 2018c]{raimbault2018unveiling}
Raimbault, J. (2018c).
\newblock Unveiling co-evolutionary patterns in systems of cities: a systematic
  exploration of the simpopnet model.
\newblock {\em arXiv preprint arXiv:1809.00861}.

\bibitem[Raimbault, 2019]{raimbault2018modeling}
Raimbault, J. (2019).
\newblock Modeling the co-evolution of cities and networks.
\newblock {\em arXiv preprint arXiv:1804.09430}.

\bibitem[Raimbault et~al., 2019]{raimbault2019}
Raimbault, J., Cottineau, C., Le~Texier, M., Le~Nechet, F., and Reuillon, R.
  (2019).
\newblock Space matters: Extending sensitivity analysis to initial spatial
  conditions in geosimulation models.
\newblock {\em Journal of Artificial Societies and Social Simulation},
  22(4):10.

\bibitem[Raimbault et~al., 2018]{raimbault2018space}
Raimbault, J., Cottineau, C., Texier, M.~L., N{\'e}chet, F.~L., and Reuillon,
  R. (2018).
\newblock Space matters: extending sensitivity analysis to initial spatial
  conditions in geosimulation models.
\newblock {\em arXiv preprint arXiv:1812.06008}.

\bibitem[Raimbault and Perret, 2019]{raimbault2019generating}
Raimbault, J. and Perret, J. (2019).
\newblock Generating urban morphologies at large scales.
\newblock {\em arXiv preprint arXiv:1903.06807}.

\bibitem[Rakshit et~al., 2017]{rakshit2017noisy}
Rakshit, P., Konar, A., and Das, S. (2017).
\newblock Noisy evolutionary optimization algorithms--a comprehensive survey.
\newblock {\em Swarm and Evolutionary Computation}, 33:18--45.

\bibitem[Reuillon et~al., 2010]{reuillon2010declarative}
Reuillon, R., Chuffart, F., Leclaire, M., Faure, T., Dumoulin, N., and Hill, D.
  (2010).
\newblock Declarative task delegation in openmole.
\newblock In {\em 2010 International Conference on High Performance Computing
  \& Simulation}, pages 55--62. IEEE.

\bibitem[Reuillon et~al., 2013]{reuillon2013openmole}
Reuillon, R., Leclaire, M., and Rey-Coyrehourcq, S. (2013).
\newblock Openmole, a workflow engine specifically tailored for the distributed
  exploration of simulation models.
\newblock {\em Future Generation Computer Systems}, 29(8):1981--1990.

\bibitem[Reuillon et~al., 2015]{reuillon2015new}
Reuillon, R., Schmitt, C., De~Aldama, R., and Mouret, J.-B. (2015).
\newblock A new method to evaluate simulation models: The calibration profile
  (cp) algorithm.
\newblock {\em Journal of Artificial Societies and Social Simulation},
  18(1):12.

\bibitem[Rey-Coyrehourcq, 2015]{rey2015plateforme}
Rey-Coyrehourcq, S. (2015).
\newblock {\em Une plateforme int{\'e}gr{\'e}e pour la construction et
  l'{\'e}valuation de mod{\`e}les de simulation en g{\'e}ographie}.
\newblock PhD thesis, Paris 1-Panth{\'e}on-Sorbonne.

\bibitem[Reymond, 1971]{reymond1971pour}
Reymond, H. (1971).
\newblock Pour une problématique théorique.
\newblock In Isnard, H., Racine, J.~B., and Reymond, H., editors, {\em
  Probl{\'e}matiques de la g{\'e}ographie}, volume~29. Presses Universitaires
  de France-PUF.

\bibitem[Rist, 1970]{rist1970student}
Rist, R. (1970).
\newblock Student social class and teacher expectations: The self-fulfilling
  prophecy in ghetto education.
\newblock {\em Harvard educational review}, 40(3):411--451.

\bibitem[Rozenfeld et~al., 2008]{rozenfeld2008laws}
Rozenfeld, H.~D., Rybski, D., Andrade, J.~S., Batty, M., Stanley, H.~E., and
  Makse, H.~A. (2008).
\newblock Laws of population growth.
\newblock {\em Proceedings of the National Academy of Sciences},
  105(48):18702--18707.

\bibitem[Sanders et~al., 2007]{sanders2007artificial}
Sanders, L., Favaro, J., Glisse, B., Mathian, H., and Pumain, D. (2007).
\newblock Artificial intelligence and collective agents: the eurosim model.
\newblock {\em Cybergeo}, 392:15.

\bibitem[Sanders et~al., 1997]{sanders1997simpop}
Sanders, L., Pumain, D., Mathian, H., Gu{\'e}rin-Pace, F., and Bura, S. (1997).
\newblock Simpop: a multiagent system for the study of urbanism.
\newblock {\em Environment and Planning B: Planning and design},
  24(2):287--305.

\bibitem[Schmitt, 2014]{schmitt2014modelisation}
Schmitt, C. (2014).
\newblock {\em Mod{\'e}lisation de la dynamique des syst{\`e}mes de peuplement:
  de SimpopLocal {\`a} SimpopNet}.
\newblock PhD thesis, Universit{\'e} Panth{\'e}on-Sorbonne-Paris I.

\bibitem[Schmitt et~al., 2017]{schmitt2017simpoplocal}
Schmitt, C., Rey-Coyrehourcq, S., Pumain, D., and Reuillon, R. (2017).
\newblock The simpoplocal model.
\newblock In {\em Urban Dynamics and Simulation Models}, pages 21--35.
  Springer.

\bibitem[Schmitt et~al., 2015]{schmitt2015half}
Schmitt, C., Rey-Coyrehourcq, S., Reuillon, R., and Pumain, D. (2015).
\newblock Half a billion simulations: Evolutionary algorithms and distributed
  computing for calibrating the simpoplocal geographical model.
\newblock {\em Environment and Planning B: Planning and Design},
  42(2):300--315.

\bibitem[Sharma and Singh, 2006]{sharma2006multi}
Sharma, S. and Singh, H. (2006).
\newblock Multi-agent system for simulating human behavior in egress
  simulations.
\newblock In {\em Proceedings of NAACSOS, Annual Conference of the North
  American Association for computational social and organizational sciences,
  Notre Dame, Indiana}.

\bibitem[Sloane, 2008]{sloane2008experiences}
Sloane, T. (2008).
\newblock Experiences with domain-specific language embedding in scala.
\newblock In {\em Domain-Specific Program Development}, page~7.

\bibitem[Solomatine et~al., 1999]{solomatine1999automatic}
Solomatine, D., Dibike, Y., and Kukuric, N. (1999).
\newblock Automatic calibration of groundwater models using global optimization
  techniques.
\newblock {\em Hydrological Sciences Journal}, 44(6):879--894.

\bibitem[Stonedahl and Wilensky, 2010]{stonedahl2010evolutionary}
Stonedahl, F. and Wilensky, U. (2010).
\newblock Evolutionary robustness checking in the artificial anasazi model.
\newblock In {\em 2010 AAAI Fall Symposium Series}.

\bibitem[Stonedahl, 2011]{stonedahl2011genetic}
Stonedahl, F.~J. (2011).
\newblock {\em Genetic algorithms for the exploration of parameter spaces in
  agent-based models}.
\newblock PhD thesis, Northwestern University.

\bibitem[Terrier, 1980]{terrier1980mirabelle}
Terrier, C. (1980).
\newblock Mirabelle.
\newblock {\em Courrier des Statistiques}, 73.

\bibitem[Tisue and Wilensky, 2004]{tisue2004netlogo}
Tisue, S. and Wilensky, U. (2004).
\newblock Netlogo: A simple environment for modeling complexity.
\newblock In {\em International conference on complex systems}, volume~21,
  pages 16--21. Boston, MA.

\bibitem[Van~Deursen and Klint, 2002]{van2002domain}
Van~Deursen, A. and Klint, P. (2002).
\newblock Domain-specific language design requires feature descriptions.
\newblock {\em Journal of Computing and Information Technology}, 10(1):1--17.

\bibitem[White and Engelen, 1993]{white1993cellular}
White, R. and Engelen, G. (1993).
\newblock Cellular automata and fractal urban form: a cellular modelling
  approach to the evolution of urban land-use patterns.
\newblock {\em Environment and planning A}, 25(8):1175--1199.

\bibitem[White et~al., 2015]{white2015modeling}
White, R., Engelen, G., and Uljee, I. (2015).
\newblock {\em Modeling cities and regions as complex systems: From theory to
  planning applications}.
\newblock MIT Press.

\bibitem[Wilson, 2014]{wilson2014complex}
Wilson, A.~G. (2014).
\newblock {\em Complex spatial systems: the modelling foundations of urban and
  regional analysis}.
\newblock Routledge.

\end{thebibliography}

\end{document}